\documentclass[flushrt,manuscript]{aastex}

\newcommand{\el}{{\rm e}}
\newcommand{\xel}{x_{\rm e}}

\newcommand{\p}{{\rm p}}
\newcommand{\h}{{\rm H}}

\newcommand{\xh}{x({\rm H})}
\newcommand{\nh}{n_{\rm H}}

\newcommand{\hp}{{\rm H}^+}

\newcommand{\he}{{\rm He}}

\newcommand{\hep}{{\rm He}^+}

\newcommand{\hetwop}{{\rm He}^{2+}}

\newcommand{\hehp}{{\rm HeH}^+}

\newcommand{\hm}{{\rm H}_2}
\newcommand{\xhm}{x({\rm H}_2)}

\newcommand{\hmp}{{\rm H}_{2}^{+}}

\newcommand{\hthreep}{{\rm H}_{3}^{+}}

\newcommand{\cp}{{\rm C}^+}

\newcommand{\xo}{x({\rm O})}

\newcommand{\oh}{{\rm OH}}

\newcommand{\ohp}{{\rm OH}^+}

\newcommand{\htwoo}{{\rm H}_{2}{\rm O}}
\newcommand{\xhtwoo}{x({\rm H}_{2}{\rm O})}

\newcommand{\htwoop}{{\rm H}_{2}{\rm O}^{+}}

\newcommand{\hthreeop}{{\rm H}_{3}{\rm O}^{+}}

\newcommand{\co}{{\rm CO}}
\newcommand{\xco}{x({\rm CO})}

\newcommand{\hcop}{{\rm HCO}^+}

\newcommand{\pcc}{{\rm cm}^{-3}}
\newcommand{\percc}{\rm \,cm^{-3}}
\newcommand{\psqcm}{{\rm cm}^{-2}}

\newcommand{\ps}{{\rm s}^{-1}}
\newcommand{\ccps}{{\rm cm}^{3} {\rm s}^{-1}}

\newcommand{\bcen}{\begin{center}}
\newcommand{\ecen}{\end{center}}
\newcommand{\be}{\begin{equation}}
\newcommand{\ee}{\end{equation}}
\newcommand{\bdis}{\begin{displaymath}}
\newcommand{\edis}{\end{displaymath}}

\def\ra{{\rightarrow}}

\shorttitle{Cosmic-ray and X-ray Heating}
\shortauthors{Glassgold et al.}

\received{2012 May 14}
\accepted{2012 July 20}

\begin{document}

\title{Cosmic-ray and X-ray Heating of Interstellar Clouds and Protoplanetary Disks}

\author{Alfred E. Glassgold}
\affil{Astronomy Department, University of California,
Berkeley, CA 94720-3411, USA}
\email{aglassgold@berkeley.edu}

\author{Daniele Galli}
\affil{INAF-Osservatorio Astrofisico di Arcetri, Largo E. Fermi 5, 50125, Italy}
\email{galli@arcetri.astro.it}

\and

\author{Marco Padovani}
\affil{Laboratoire de Radioastronomie Millim\'etrique, UMR 8112 du CNRS, \'Ecole Normale Sup\'erieure et Observatoire de Paris, 24 rue Lhomond, 75231 Paris cedex 05} 
\email{marco.padovani@lra.ens.fr}

\begin{abstract}
Cosmic-ray and X-ray heating are derived from the electron energy loss calculations of Dalgarno, Yan and Liu for hydrogen-helium gas mixtures.
These authors treated the heating from elastic scattering and collisional 
de-excitation of rotationally excited hydrogen molecules. Here we consider the heating that can arise from all ionization and excitation processes, with particular emphasis on the reactions of cosmic-ray and X-ray generated ions with the heavy neutral species, which we refer to as chemical heating.
In molecular regions, chemical heating dominates and can account for 50\% of the energy expended in the creation of an ion pair. The heating per ion pair ranges in the limit of negligible electron fraction from $\sim 4.3 $\,eV for diffuse atomic gas, to $\sim 13$\,eV for the moderately dense regions of molecular clouds and to $\sim 18 $\,eV for the very dense regions of protoplanetary disks. An important general conclusion of this study is that cosmic-ray and X-ray heating depends on the physical properties of the medium, i.e., on the molecular and electron fractions, the total density of hydrogen nuclei, and to a lesser extent on the temperature. It is also noted that chemical heating, the dominant process for cosmic-ray and X-ray heating, plays a role in UV irradiated molecular gas.

\end{abstract}


\section{Introduction}

There has been a significant increase of interest in the cosmic-ray
ionization rate in both local and distant neighborhoods. Much of it
has been spurred by infrared absorption-line measurements of the
$\hthreep$ ion in diffuse and translucent interstellar clouds 
(e.g., Indriolo et al.~2007 and references therein). Indriolo \& McCall~(2012) have extended these measurements to fifty lines of sight 
dominated by translucent clouds with more than one magnitude of visual extinction, obtaining 21 detections. Their straightforward analysis 
of the observations yields values of the ionization rate for $\hm$ in the
range $\sim (2-11) \times 10^{-16}\,\ps$, significantly larger than the
range of values that had been in use for several decades\footnote{In this 
and other papers, the cosmic-ray ionization rate $\zeta_{\rm H}$ per H nucleus is used, which is smaller by a factor of two than the rate 
$\zeta_{{\rm H}_2}$ per $\hm$ molecule of Indriolo et al.\,(2007).}. This result has led to the reconsideration of how both electron and proton 
cosmic-rays propagate in the interstellar medium (e.g., Indriolo et 
al.~2009, Padovani et al.~2009, Padovani \& Galli 2011, Rimmer et 
al.~2012, Everett \& Zweibel 2011). In the context of the new survey, Indriolo \& McCall al.~(2012) reviewed the explanation advanced by Indriolo et al.~(2009) and Padovani et al.~(2009) that the intensity of 
low-energy protons is reduced inside dense molecular clouds. The large difference between diffuse/translucent and dense clouds is maintained in Indriolo \& McCall~(2012), but their sample shows little evidence for a dependence of ionization rate on column density. The value of the cosmic
ray ionization rate for the local interstellar medium remains an open question.

Another topic of interest relating to the interaction of cosmic-rays
with the interstellar medium is their role in heating. Cosmic-rays are
an efficient (often dominant) source of heating in various
environments, from the dense gas in molecular clouds (Goldsmith \&
Langer~1978), both in normal and starbust galaxies (e.g., Suchkov et
al.~1993), to photodissociation regions (Shaw et al.~2009), and
possibly even in the primordial gas (Jasche et al.~2007). In dense,
shielded regions like molecular cloud cores, the balance of cosmic-ray
heating and gas cooling determines a temperature gradient decreasing
inwards (Galli et al.~2002) that has been accurately traced by
interferometric observations of molecular emission in prestellar cores
(Crapsi et al.~2007, Pagani et al.~2007) and dark globules (Pineda \&
Bensch~2007). The values for the cosmic-ray heating energy per 
ion pair available in the literature range over a factor of three. In 
this paper we will attempt to clarify this situation for both cosmic-ray 
and X-ray heating. 

The interaction of fast electrons slowing down in molecular gas was
first analyzed more than three decades ago by Glassgold \&
Langer~(1973, hereafter GL73), Cravens et al.~(1975) and Cravens \&
Dalgarno~(1978).  Although GL73 was the first paper to consider cosmic
ray heating in interstellar molecular regions, it suffered from the
incompletely known electron cross sections of the early 70s, a crude
energy-loss calculation, and cumbersome notation. Cravens et al.~(1975) 
did a better job on cross sections and energy loss, but restricted 
themselves to low-energy electrons in order to evaluate the
validity of the commonly used continuous slowing down approximation.
This limitation was removed by Cravens \& Dalgarno (1978). There are no
glaring discrepancies between  Cravens \& Dalgarno (1978) and GL73. All
of these papers considered pure $\hm$ regions, i.e., the roles of $\hp$
and $\hep$ ions in molecular gas were ignored. A more up to date and
complete analysis was carried out by Dalgarno et al.~(1999, hereafter
DYL). DYL considered carefully all of the energy loss channels for
electron energies from 30~eV to 1~keV in various mixtures of $\h$, $\hm$ and
$\he$. They showed how the energy expended to make  an ion pair $W$ is
partitioned between elastic and non-elastic processes, but they
did not fully treat the heating.

Cosmic-rays and X-rays (or the equivalent, fast electrons) produce
ions and excited molecules that can interact with the dominant neutral
atomic or molecular gas. The products of these reactions carry away a
significant amount of the available energy and heat the gas.  We refer
to this as chemical heating, and quantify it by a quantity $Q_{\rm chem}$
defined per ion pair. Chemical heating by cosmic-rays and X-rays
occurs in a wide range of applications from diffuse and dense
interstellar clouds to pre-stellar cores, protoplanetary disks, and
planetary atmospheres. In this paper we use the results of DYL
to inquire how the excited species and ions heat a gas
mixture of $\h$, $\hm$, and $\he$.  We will show that $\sim 1/2$ of the
energy of X-rays and cosmic rays can go into heating in dense molecular regions
and that chemical heating can be the most important part of the heating. 
We also note that chemical heating occurs when
molecular regions are irradiated by far ultra-violet (FUV) photons.

This paper is organized as follows. In Section~\ref{phys} we outline the
physical basis of our analysis and make quantitative estimates in
Sections~\ref{heating} and~\ref{chem}. We discuss the most important results in
Section~\ref{results}.  The paper ends with a short summary in
Section~\ref{conc}.

\section{The Physical Basis for Cosmic-ray and X-ray Heating}
\label{phys}

A basic tenet of earlier work on cosmic-ray heating is that 
proton cross sections can be represented accurately by the 
corresponding ones for electrons (which are generally more
available), at least at sufficiently large energy. Cravens 
\& Dalgarno (1978) discussed this explicitly on the basis of 
the good agreement between experiments on proton collisional 
ionization of $\hm$ (Hooper et al.~1961) and the Born approximation, 
as developed by Bates \& Griffing (1953), all the way down to 
a proton energy of 0.06\,MeV. This conclusion stems from the 
fact that high-energy ionization cross sections are a function 
of the incident particle's speed, so proton cross sections are 
the same as those for electrons with the same speed, i.e., 
when the electron energy $E$ satisfies the relation 
$E=(m_{\el}/m_{\p})E_{\p}$. 

An energetic cosmic-ray proton loses energy in cosmic gases in 
small steps that are characterized by the energy $W$ to make an 
ion pair, known to be close to 37\,eV from extensive experimental and 
theoretical studies. In their interactions with the hydrogen and 
helium of the gas, each cosmic-ray (proton) ionization generates a 
distribution of relatively slow electrons with average energy of 
19\,eV (Opal et al.~1971; GL73) and that leads to another ``secondary'' ionization two-thirds of the time.
X-rays with keV energies develop a similar sequence once the initial
photon is absorbed by a heavy atom by L- or K-shell electron ejection. 
The energy of this primary electron is $E=h\nu_X -B - {\cal E}$,
where $B$ is the binding energy of the K or L shell of the heavy atom and
${\cal E}$ is the mean energy of the resulting ion. It is significantly 
less than the initial X-ray energy, but much of the difference is 
restored in the form of several prompt supra-thermal Auger electrons 
with keV energies. There is also some X-ray fluorescence that, by the processes just described, generates more fast electrons. The net effect 
of X-ray ionization of heavy atoms, accompanied by Auger electrons and fluorescence, is that most of the initial X-ray photon energy is recovered 
in the form of the primary electron plus several other fast electrons. Thus, once 
the the processes attending the absorption of a keV X-ray have occurred, 
the main ionization effects of the X-ray are accomplished by the fast 
electrons that it produces. This is almost exactly the situation with
cosmic-ray ionization, except that these electrons are replaced by the 
cosmic-ray proton which, as noted above, behaves like an electron of the same 
speed. This is the basis for our using the extensive results of DYL 
on the interaction of electrons with atomic and molecular gases 
to calculate the heating by cosmic-rays and X-rays. 

The spectra of cosmic-rays and X-rays can change from region to region 
due to proximity to a source or absorption within the cloud of interest.
One example is the absorption of low-energy protons in thick clouds, 
proposed by Indriolo et al.~(2009) and Padovani et al.~(2009) to explain  the reduction of the ionization rate inside dense molecular clouds. Another is the 
observed variation in the observed spectra of the X-rays emitted by young stellar objects. DYL give results for mono-energetic electrons with energies from 30 to 1000\,eV. Their results do not vary by more than 10\% for $E > 200$\,eV, and if the underlying spectrum does not extend below this energy for electrons or the equivalent energy of 0.4\,Mev for protons, the variation of the DYL results at low electron energy can be ignored. Otherwise the results presented in this paper have to be integrated over the appropriate spectrum. 
Below $E = 30$\,eV (or $\sim 60$\,keV protons), the lowest energy considered 
by DYL, one can use GL73, but the equivalence between electrons and protons has very likely broken down at this point. In this paper, we focus on electrons with energies greater than 500\,eV, corresponding to protons with 
energies greater than 1\,MeV.  
   
We use DYL to analyze all of the outcomes of a fast electron as it
slows down by interacting with a cosmic gas-mixture dominated by $\h$,
$\hm$, and $\he$, ignoring the much less abundant heavy elements. Unlike
DYL, we make a comprehensive effort to evaluate how much of the energy
expended in each process goes into heating -- the primary goal of this
paper. For this purpose, the most useful results in DYL are Table~4
(ion production), Table~5 (excitation), and Table~7 (heating
efficiency), all for the case of H, $\he$ and $\hm$, $\he$ gas
mixtures.

\begin{deluxetable}{lrrrr}
\tabletypesize{\scriptsize}
\tablewidth{0pt}
\tablecaption{DYL Energy Partitions for a  $\he$, $\hm$ mixture$^{\rm a}$}  
\tablehead{
\colhead{Process} & \colhead{$W$} & \colhead{$N$} & \colhead{$E_{\rm th}$} & \colhead{$NE_{\rm th}$}   \\        
 & \colhead{(eV)} & & \colhead{(eV)} & \colhead{(eV)}
}        
\startdata
$\hmp$			& 41.9	 & 23.87	& 15.44	& 369	\\
$\hp$			& 921	 & 1.09	& 18.08	&  20	\\
$\hep$			& 459	 & 2.18	& 24.59	&  54	\\
He$^{2+}$		& 16000 & 0.0625	& 54.42	&   3	\\
\tableline
\tableline
Total Ion production	&	 &		&		& 446	\\
\tableline
B			& 117	 & 8.55		& 11.37		&  97	\\
C			& 132	 & 7.58		& 12.42		&  94	\\
Dissociation$^{\rm b}$	& 92.6	 & 10.8		& $\sim 13$   & 140	\\
H+H($2p$)		& 534	 & 1.87		& 14.7		&  28	\\
$v=1$			& 7.81	 & 128		& 0.516		&  66	\\
$v=2$			& 109	 & 9.17		& 1.032		&  10 	\\
\tableline
\tableline
Total $\hm$ excitation&	 &		&		& 435	\\
\tableline
He 2$^1$S\,\,$^{\rm c}$	& 15000 & 0.0667	& 20.62		&   1	\\
2$^1$P			& 1940	 & 0.515	& 21.22		&  11	\\
2$^3$S			& 34900 & 0.0287	& 19.82		&   1	\\
2$^3$P + $n>2^3$	& 22500	 & 0.0444	& 21.0		&   1	\\
$n>2^1$			& 3520	 & 0.284	& 23		&   7	\\
\tableline
\tableline
Total He excitation	&	 &		&		&  21	\\
\tableline
Elastic collisions and		&&&			&  	\\
Rotational excitation$^{\rm d}$ 		&&&			&  57	\\
\tableline
\tableline
Grand total		&	 &		&		& 959 \\
\enddata
\tablenotetext{a}{For the case $\xel = 0$ and $E=1$\,keV}
\tablenotetext{b}{Dissociation occurs mainly via triplet excitations.}
\tablenotetext{c}{He excitation follows the notation in DYL.}
\tablenotetext{d}{Based on DYL ``heating efficiency'' in their Table 7.}
\end{deluxetable}

The energy partitions for a 1-keV electron in a $\he$, $\hm$ mixture
in the limit of zero electron fraction are listed in Table 1. 
The entries are independent of energy above 100--200 eV. If the X-ray 
spectrum extends below 200 eV or the cosmic-ray spectrum below 400 keV, the 
full tables in DYL contain the information needed for lower energies. 
Table 1 has been organized into four parts: ion production; excitation 
of $\hm$ levels; excitation of $\he$ levels; and the heating from 
elastic scattering combined with the heating from rotational excitation. 
All energies are expressed in eV. The first column names the process; 
the second column the energy expended for that process $W$; the third 
column the number of such processes for an electron energy $E$, $N = E/W$,
whe the electron energy is $E=1$\,keV; 
the fourth column the threshold energy $E_{\rm th}$ for the process; 
and lastly the energy associated with the process, i.e., the product 
$NE_{\rm th}$ of the two previous columns (for $E=1$\,keV).  

For example the first four lines of Table 1 show that a keV electron 
generates 23.9 $\hmp$ ions, 1.1 $\hp$ ions, 2.2 $\hep$ ions, and 
$< 0.1$ He$^{2+}$ ions, for a total of 27.2 ions. This follows from 
the fact that an incident electron expends on average an energy $W_i$ to make 
an ion pair of type $i$ ($i=1$ for $\hmp$, $i=2$ for $\hp$, $i=3$ 
for $\hep$, and $i = 4$ for He$^{2+}$).  More generally, if $N_i = E/W_i$ 
is the number of ions generated of type $i$, the total number of ions is 
$N = \sum_i N_i = E \sum_i W_i^{-1}$ and the average energy per ion pair is
\be
\frac{E}{N} = (\sum_i W_i^{-1})^{-1}\,.
\ee 
According to Table 1, $E/N= 36.8$~eV per ion pair for $E=1$\,keV 
and zero electron fraction.

The data in DYL account for 96\% of the initial electron energy  $E=1$\,keV 
in Table 1. The small deviation from 100\%
is not significant in light of the uncertainties in the calculations. It may 
be due to the use of the approximate fitting formulae for the parameters $W$, 
for which DYL quote an accuracy of 5-15\%. Another possibility is the omission 
of triplet states starting near 14\,eV that lie above the 
$a$, $b$, and $c$ levels, which DYL refer to as missing ``pseudo states''. In addition, 
most of the cross sections used in the energy loss process are probably not 
known to an accuracy of 10\%.  Approximately 47\% of the energy in Table 1 
is accounted for by ionization, 47\% by excitation, and 6\% by direct heating and
the collisional de-excitation of $\hm$ rotational levels. Applying this to 
the 36.8\,eV to make an ion pair, elastic scattering and rotational excitation lead 
to only about 2\,eV heating per ion pair in neutral gas. However, as the ionization 
fraction increases beyond $\xel = 10^{-4}$, the direct heating increases, largely 
due to Coulomb  collisions with ambient electrons\footnote{It is important 
to note that the DYL considerations do not apply above $\xel = 0.1$.}. 
Most cosmic-ray and X-ray heating in molecular regions comes from the 
reactive ions and excited states. This is exactly the part of the energy loss 
problem that is {\it not} treated by DYL. 

\section{The Heating Energy}
\label{heating}
In general the heating energy per ion pair $Q$ consists of contributions from
collisions with H and $\hm$, 
\be Q = \frac{x({\h})Q({\h}) +
x(\hm)Q(\hm)}{x({\h}) + x(\hm)}\,, 
\label{defQ} 
\ee 
where $\xh+2\xhm=1$;
the functions $Q(\h)$ and $Q(\hm)$ depend on the abundance of $\he$ and
especially on the abundance of electrons. The volumetric rate of ion-pair
production is $\zeta_{\rm H} \nh$, where $\zeta_{\rm H}$ is the ionization 
rate per H nucleus and $n_{\rm H}$ is the volumetric density of H nuclei. 
Thus the volumetric heating rate is given by 
$\Gamma = Q \, \zeta_{\rm H} n_{\rm H}$.

The dominant heating process in atomic regions is elastic scattering.
DYL include Coulomb collisions with ambient electrons, but only for 
the case where the electron abundance relative to total $\h$ nuclei 
is $< 0.1$. For negligible ionization, it is twice that for $\hm$ 
regions, as can be seen in Table A of the Appendix, the analog of 
Table 1 for atomic regions. It is included in $Q({\h})$, the first 
term in the numerator of Eq.~\ref{defQ}. In Appendix A we show that 
chemical heating induced by the atomic ions $\hp$ and $\hep$ reacting 
with $\h$ is very small. As shown in DYL Figure 8, the dominant 
excitation product in atomic H is the $2p$ level of atomic hydrogen. 
The resultant Ly-$\alpha$ photons will either escape or be absorbed 
by and heat the dust, and not the gas. 

The heating in molecular regions is the sum of the effects from 
elastic collisions plus rotational excitation ($Q_{\rm el/rot}$), excitation of 
$\hm$ vibrational levels ($Q_{\rm vib}$), dissociation of $\hm$ ($Q_{\rm diss}$), 
and chemical heating ($Q_{\rm chem}$):
\be
\label{totQ}
Q(\hm) = Q_{\rm el/rot} + Q_{\rm vib} + Q_{\rm diss} + Q_{\rm chem}\,.
\label{defQH2}
\ee
In the following paragraphs we discuss the first three contributions to $Q(\hm)$, 
and then focus on $Q_{\rm chem}$ in the next section.

\subsection{Elastic Scattering and Rotational Excitation}

The critical densities for the rotational levels of $\hm$ depend on the transition 
and also on the H/$\hm$ ratio and the temperature (e.g., Le Bourlot et al.~1999). In 
many situations the lowest S(0) and S(1) rotational levels are collisionally 
de-excited and most of the rotational excitation goes into heating. GL73 estimated that a density $\nh = 1000 \percc$ would satisfy this condition on density, independent of temperature. We use  DYL for the combined heating due to elastic scattering and rotational excitation, 
\be
Q_{{\rm el/rot}}(\hm,\el) \equiv Q_{{\rm el}}(\hm,\el) + Q_{{\rm rot}}(\hm,\el)\,. 
\ee
This quantity is then expressed in terms of a ``heating efficiency'' $\eta$ for a $\hm$, 
He gas mixture,
\be
Q_{{\rm el/rot}}(\hm,\el) = \eta(\hm,\el) W(\hm,\el)\,,
\ee
where $W(\hm,\el)$ is the average energy to make an ion pair. Both $\zeta_{\rm H}$ and the
heating function $Q$ are defined in terms of ion pairs. For 
the H, He gas, there is of course no rotational heating, and the heating is expressed 
in a similar form where the heating efficiency $\eta(\h,\el)$ now describes only 
elastic collisions, 
\be
Q_{{\rm el}}(\h,\el) = \eta(\h,\el) W(\h,\el)\,.
\ee
However, $W(\h,\el) \approx W(\hm,\el)$ to an accuracy of $\approx 5\%$ for electron 
energies greater than several hundred eV, and the first term of 
Eq.~\ref{defQH2} becomes,
\be
\label{heating_efficiency}
Q_{{\rm el}}(\h,\el) + Q_{{\rm el/rot}}(\hm,\el) \approx 
\frac{[\xh \eta(\h,\el) + \xhm \eta (\hm,\el)]}{ \xh + \xhm}\, W(\hm,\el)
\ee 
with $W(\hm, \el)\approx 37$\,eV asymptotically for $E \geq 500$\,eV. 

The heating efficiencies $\eta$ are expressed by the fitting formulae, DYL Eq.~(14),
\be
\eta = 1 - \frac{1-\eta_0}{1 + C \xel^{\alpha}}\,.
\ee
The two sets of fit parameters are given at the bottom of DYL Table 7 
("Two-Gas Mixtures") as a function of electron energy $E$. For example, 
the $E = 1$\,keV values are, 
\be
\eta(\h,\el)_0 = 0.117, \hspace{0.25in} \alpha = 0.678, \hspace{0.25in} C = 7.95,
\ee
for atomic regions, and 
\be
\eta (\hm,\el)_0 = 0.055, \hspace{0.25in} \alpha = 0.366, \hspace{0.25in} C = 2.17,
\ee
for molecular regions. These values might be appropriate for the keV X-rays emitted 
by young stellar objects. These fits have only a modest accuracy $\sim 10-15$\,\%.

According to DYL (see Table 1),  $Q_{\rm el/rot}$
accounts for only 57~eV, or $\sim 6$\% of the energy of the 1-keV
incident electron. The rest of the energy is almost equally shared by
ionization and excitation of $\he$ and $\hm$.  Thus, the DYL heating 
efficiency for molecular regions is only a small part of the total.  
With $W \sim 37$~eV to make an ion pair, elastic scattering and rotational 
excitation lead to $Q_{\rm el/rot}=2.1$~eV heating per ion pair in neutral 
$\hm$, $\he$ gas. This is half of the direct heating in an atomic $\h$, $\he$, 
$\sim 4.3$~eV under the same conditions.  

\subsection{Dissociation Heating}

The most important pathway to dissociation is collisional excitation of $\hm$ 
{\rm triplet} states that start at 11.9\,eV ({\it a} and {\it c}), 13.4\,eV 
({\it e}) and possibly above. These levels decay into continuum states 
{\it b} of the triplet repulsive potential and have essentially unit probability for dissociation. Unlike the Lyman and Werner transitions to the singlet $B$ and $C$ levels, these spin-flip  transitions are forbidden for photon excitation from the singlet ground state $X$. Table 1 indicates that for $E=1$\, keV and $\xel =0$, there are 10.8 dissociations per keV. Adopting 
the DYL estimate of 5.4\,eV per dissociation, we obtain 58\,eV heating per keV, 
or 2.14\,eV per ion pair for a neutral $\hm$, He mixture \footnote{This number does not include the heating 
from dissociation following collisional-excitation of the $B$ and $C$ states, 
which is much smaller, only $\sim 0.6$\,eV per keV.}. More generally we use 
DYL Table 5 for the  dissociation heating in Eq.~\ref{totQ},
\be
Q_{{\rm diss}} = \frac{\xhm}{\xh + \xhm}\,\frac{D_0}{1+ C \xel ^{\alpha}}\,,
\ee 
where $D_0 =2.14$\,eV, $\alpha = 0.574$ and $C = 22.0$. 

\subsection{Vibrational Heating}

Fast electrons can excite the vibrational levels of $\hm$ by direct collisions 
or by fluorescent de-excitation of electronically excited $B$ and $C$ levels.  
The excitation energy goes into heating if the densities are high enough for the levels to be collisionally de-excited. Recent calculations of the collisional de-excitation rate coefficients (e.g. Wrathmall et al.~2007 for H; Lee et al.~2008 for $\hm$; Balakrishnan et al.~1999 for He) replace earlier work of Tin\'e et al.~(1997) and Le Bourlot et al.~(1999). 
The coefficients are generally much larger for collisions with atomic H than for $\hm$ and He.  They are very small below 100\,K  and increase rapidly above this temperature. Thus one can expect little quenching of vibrational levels in the cool and moderately dense regions of molecular clouds. 
By contrast, in the surface layers of the inner regions of protoplanetary disks, where $\nh \sim 10^{10} \pcc$ and $T \sim 1000$\,K, the densities of both H and $\hm$ are large enough to collisionally quench vibrationally-excited $\hm$ molecules and provide significant heating. 
In less dense molecular regions, atomic H collisions may be effective in collisionally de-exciting $\hm$, but this depends on the temperature and the $\h/\hm$ ratio as well as the total density $\nh$. 

In order to obtain the heating from direct collisional excitation in dense $\hm$
regions, we use DYL Table 5 for $\hm$, He mixtures which gives the following 
fit for the energy to excite the $v=1,2$ levels,
\be
\label{vib_paras}
\epsilon_v = W_v (1 + C_v\, \xel^{\alpha_v})\,. 
\ee
The parameters for $E=1$\,keV are given in Table 2 for $E \geq 500$\,eV, 
with $W_v$ and $ \Delta E_v$ (the vibrational excitation energies) in eV.

\begin{deluxetable}{lcccc}
\tablewidth{0pt}
\tablecaption{Direct Vibrational Parameters$^{\rm a}$} 
\tablehead{
\colhead{$v$} & \colhead{$W_v$ (eV)}	& \colhead{$C_v$} & \colhead{$\alpha_v$} 	& \colhead{$\Delta E_v$ (eV)}
}
\startdata
1	& 7.81	 & 23,500	& 0.955	& 0.516	\\
2	& 109	 & 10,700	& 0.907	& 1.032	\\	
\enddata
\tablenotetext{a}{$E=1$\,keV}
\end{deluxetable}
 
Because $1000/\epsilon_v$ is the number of excitations per keV of electron energy, 
the high-density direct vibrational heating per keV of incident electron energy is, 
\[
\left(\frac{1000}{\epsilon_1}\,\Delta E_1 + 
\frac{1000}{\epsilon_2}\,\Delta E_2\right)\;{\rm eV}\,.
\]
The vibrational heating per ion pair is obtained by dividing this expression 
by the number of ion pairs per keV (27.2 for $E \geq 500$\,eV), with the result,
\be
\label{qdirvib}
Q_{\rm dir/vib} \approx \frac{\xhm}{\xh + \xhm} \,
19.0\,{\rm eV}\, 
\left[ \, \frac{1}{\epsilon_1} + \frac{2}{\epsilon_2} \, \right]\,.
\ee
In a pure $\hm$ region with negligible ionization, the vibrational heating per ion pair 
is 2.8\,eV for $E \geq 500$\,eV, with $v=2$ contributing about 15\%.

Excited vibrational levels are also produced when collisionally excited
$B$ and $C$ states decay with the emission of fluorescent photons in the 
1500--1600~\AA\, range, with a maximum near 1575~\AA. We can calculate the 
high-density heating that follows the collisional de-excitation of the  
vibrational levels using the above method for direct collisional excitation 
based on Eq.~\ref{vib_paras}. The necessary parameters again come from 
DYL Table 5 for $\hm$, He mixtures.

\begin{deluxetable}{lccc}
\tablewidth{0pt}
\tablecaption{$B$, $C$ Vibrational Parameters$^{\rm a}$} 
\tablehead{
\colhead{Level} & \colhead{$W_v$ (eV)}	& \colhead{$C_v$} & \colhead{$\alpha_v$}
}
\startdata
$B$	& 117  & 7.09	& 0.779	\\
$C$	& 132	 & 6.88	& 0.802	\\	
\enddata
\tablenotetext{a}{$E=1$\,keV}
\end{deluxetable}

With roughly 4\,eV in vibrational excitation energy and 27.2 ion pairs 
per keV, the high-density $B$, $C$ vibrational heating per keV of incident electron energy 
is,
\[
\frac{ 1000 / \epsilon_B + 1000 / \epsilon_C}{27.2}
\,4.0\, {\rm eV} = 147\,{\rm eV}\left[ \frac{1}{\epsilon_B} + \frac{1}{\epsilon_C}\right]\,.
\]
The $B$, $C$ vibrational heating per ion pair is then,
\be
\label{qBCvib}
Q_{BC/{\rm vib}} = \frac{\xhm}{\xh + \xhm} \, 147\,{\rm eV} \, 
\left[ \, \frac{1}{\epsilon_B} + \frac{1}{\epsilon_C} \, \right]\,.
\ee
If we substitute the values in Table 3 for $\xel = 0$, the result is 2.4\,eV
per ion pair for a neutral $\hm$, He gas, a small but not negligible contribution to the high-density heating.

As discussed earlier, both Eqs.~\ref{qdirvib} and \ref{qBCvib} hold only 
if the density well exceeds the critical density $n_{\rm cr}$ 
for the de-excitation of the vibrational transitions of the $\hm$ ground level. 
They therefore should be multiplied by an appropriate factor $\Theta_{\rm vib}$, 
that depends on the ratio $n_{\rm cr}/\nh$, such that $\Theta_{\rm vib} = 0$ for small 
$n_{\rm cr}/\nh$ and $\Theta_{\rm vib} = 1$ for large $n_{\rm cr}/\nh$. Their 
contribution to the total heating in Eq.~\ref{totQ} is now,
\be
\label{totvib}
Q_{\rm vib} = \Theta_{\rm vib} \, (Q_{\rm dir/vib}+ Q_{BC/{\rm vib}})\,.
\ee

The $B$ and $C$ levels have a $\sim 15$\% probability to dissociate 
(e.g., Abgrall et al.~1997), with a yield of approximately 0.25 eV in typical interstellar conditions (Tielens 2000).  The small branching ratio for dissociation and the small energy yield mean that the heating per ion pair following dissociation is very small, e.g., 0.02\,eV for $E=1$~keV, and thus negligible. 

\section{Chemical Heating} 
\label{chem}

Chemical heating derives mainly from reactions instigated by the primary
cosmic-ray and X-ray ions, $\hmp$, $\hp$, and $\hep$, with neutral species 
and electrons. It was first considered in the context of EUV heating of 
Jupiter's upper atmosphere (Henry \& McElroy~1969) and then for cosmic 
ray ionization of interstellar molecular clouds (GL73). These authors 
focused on $\hmp$, its transformation into $\hthreep$, 
and the destruction of $\hthreep$ by dissociative recombination. In dense 
molecular regions, however, where $\xel \ll \xhm$, dissociative recombination 
has to compete against ionic reactions with neutral species. Rather than treat 
all of the ion-neutral reactions that can be traced to the primary ions, 
$\hmp$, $\hp$, and $\hep$, we focus on only the potentially most abundant 
neutrals: $\co$, $\htwoo$, and O. Although this treatment is not completely general, it should suffice to demonstrate the nature and magnitude of chemical heating in molecular regions. The chemical heating will be expressed in terms of 
its values for the three main initiating ions in $\hm$ regions,
\be
\label{chemQsum}
Q_{\rm chem}(\hm) = Q_{\rm chem}(\hmp,\hm) + Q_{\rm chem}(\hp,\hm) +
Q_{\rm chem}(\hep,\hm)\,.
\ee
According to the discussion of Table 1 in Section~\ref{phys}, each of
these terms includes a weighting factor equal to the fraction for each
ion of the total, $F(\hmp) =0.88$, $F(\hp) = 0.04$, and
$F(\hep)=0.08$, ignoring the tiny fraction of $\hetwop$.  In Appendix
A we show that chemical heating in atomic regions is negligible.

\subsection{Chemical Heating of {\rm H}$_2$}

As discussed in Section 2, the most abundant ion generated by cosmic-rays 
and X-rays in $\hm$ regions is $\hmp$. We first consider its main 
destruction routes; $\hp$ and $\hep$ ions will be analyzed later in  
this section. $\hmp$ is mainly destroyed by dissociative recombination and 
by proton transfer with $\hm$,
\be
\label{h2p}
\hmp + \el \, \ra \, \h + \h\,, \hspace{0.5in} \hmp + \hm \, \ra \,  \hthreep + \h\,.
\ee
with rate coefficients, $\beta = 2.0 \times 10^{-7}\,T^{-1/2} \ccps$ (a rough fit to Schneider et al.~1994) and 
$k = 2\times 10^{-9} \ccps$ (Theard \& Huntress~1974). 
The probability for $\hthreep$ production is,
\be
\label{convprob}
P(\hmp,\hthreep) = \frac{k\xhm}{k\xhm + \beta \xel}=
                   \frac{\xhm}{\xhm + 100 T^{-1/2}\xel}\,,
\ee 
so that if,
\be
\label{xel_condition}
\xel <  0.1\, \frac{k}{\beta} \xhm = 0.01 \, 
\left(\frac{T}{100~{\rm K}}\right)^{1/2}\, \xhm\,, 
\ee 
the rate of dissociative recombination is less than $10\%$ that for $\hm$ 
destruction. For regions with considerable $\hm$, this condition is well 
satisfied, $P(\hmp,\hthreep) \approx 1$, and most of the $\hmp$ ions 
are transformed into $\hthreep$.

The destruction pathways for $\hthreep$ are first, dissociative recombination 
with two branches,
\be
\label{h3p_diss}
\hthreep + \el \, \ra \,  \hm + \h \; (25\%)\hspace{0.15in} {\rm and} \hspace{0.15in}
\h + \h + \h \;(75\%)\,,
\ee
and total rate coefficient $\beta' = 4.5 \times 10^{-6}\,T^{-0.65} \ccps$ (Sundstr\"om et al.~1994), 
and second, proton-transfer reactions such as,
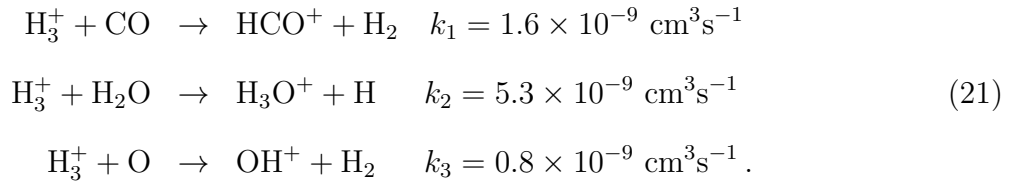
\begin{eqnarray}
\label{h3p}
\hthreep + \co     & \ra & \hcop + \hm     \hspace{0.16in} k_1 = 1.6 \times 10^{-9}~\ccps \nonumber \\
\hthreep + \htwoo  & \ra & \hthreeop + \h \hspace{0.25in} k_2 = 5.3 \times 10^{-9}~\ccps \\
\hthreep + {\rm O} & \ra & \ohp + \hm     \hspace{0.25in} k_3 = 0.8 \times 10^{-9}~\ccps \,.
\nonumber 
\end{eqnarray}
Dissociative recombination, Eq.~\ref{h3p_diss}, was considered by GL73. It is
competitive with the reactions in Eq.~\ref{h3p} where the electron fraction is 
relatively large. Denoting the abundance of the neutral reactants 
in Eq.~\ref{h3p} by $x_i$ with $i= 1, 2, 3$ for CO, $\htwoo$, and O,  
respectively, we introduce branching
ratios for the main reaction pathways of $\hthreep$,
\be
\label{bratios}
B_{\rm e} = \frac{\beta' \xel}{\Sigma_i k_ix_i+ \beta'\xel} \hspace{0.50in}
B_i = \frac{k_i \,x_i}{\Sigma_i k_ix_i+ \beta'\xel} \hspace{0.25in}(i=1,2,3) \,.
\ee
In order for dissociative recombination to compete with ionic reactions at the
10\% level, the electron fraction must satisfy the condition,
\be
\label{dr_dominance}
\xel > 0.1 \, \frac{\Sigma_i k_ix_i}{\beta'} = 
		2.22 \times 10^{-9} \, T^{0.65}\, (\Sigma_i k_ix_i)_{-13}\,,
\ee
where $(\Sigma_i k_ix_i)_{-13} = (\Sigma_i k_ix_i/ 10^{-13}~\ccps)$ 
measures the reactivity of $\hthreep$ with abundant neutrals in units
determined by neutral abundances of order $10^{-4}$ and ionic reactions 
with rate coefficients of order $10^{-9}~\ccps$. Even in cold molecular 
regions,  ($T \sim 10$\,K) dissociative recombination can still play a 
role in the presence of neutral reactions. Eq.~\ref{dr_dominance} for 
$\hthreep$ is much less restrictive than Eq.~\ref{xel_condition} for 
$\hmp$ because of the reduced abundance of heavy elements compared to 
hydrogen.

The first reaction in Eq.~\ref{h3p} with CO is simple to treat because the 
product $\hcop$ is mainly destroyed in one step by dissociative recombination 
back to CO + H, whereas dissociative recombination of $\hthreeop$ in the second 
reaction has three branches, $\htwoo$ + H, OH + $\hm$, and OH + 2H. The situation 
for the reaction with O is similar because $\ohp$ in the third equation is quickly 
transformed into $\hthreeop$ by hydrogenation reactions with $\hm$. In this 
sequence, we can ignore dissociative recombination of $\ohp$ and $\htwoop$ 
because of conditions similar to Eq.~\ref{xel_condition}.

Keeping in mind that the probability for the conversion of $\hmp$ into 
$\hthreep$, $P(\hmp,\hthreep)$, is essentially unity according to 
Eqs.~\ref{convprob} and \ref{xel_condition}, the chemical energy released 
by the creation of an $\hmp$ ion can be obtained by consolidating the reactions 
in Eq.~\ref{h2p} and Eq.~\ref{h3p} into the equivalent reactions,
\be
\label{coheat}
\el + \hmp  \;\; \ra \;\; \h+\h 
\hspace{0.20in}
\ee 
\be
\label{waterheat}
\el + \hmp + \htwoo + \hm
\;\;\ra \;\;\htwoo + \hm + 2\h\, ,\;\; \oh + 2\hm + \h, \;\; \oh + \hm + 3\h 
\ee
\be
\label{atomicOheat}
\el + \hmp + {\rm O} + 2\hm \;\; \ra\;\; \htwoo + 4\h\, ,\;\; \oh + \hm + 3\h\, , 
\;\; \oh + 5\h
\ee
where $\overline{q}_1 = 11.1\,{\rm eV}$, $\overline{q}_2 = 7.8\,{\rm eV}$, and 
$\overline{q}_3 = 5.7\,{\rm eV}$ are the net energy yields for each of 
the three consolidated reactions of $\hmp$, Eqs.~\ref{coheat}, ~\ref{waterheat} 
and ~\ref{atomicOheat}. The $\overline{q}$ are averages over the 
outcomes generated by the three branches in the  dissociative recombination of 
$\hthreeop$, 
\be
\label{h3op+branches}
\hthreeop + \el \;\; \ra \;\; \htwoo + \h \; (25\%); \hspace{0.15in}
\oh + \hm \; (15\%); \hspace{0.15in} \oh + 2\h \; (60\%) \,.
\ee
For example, Eq.~\ref{coheat} results from adding the full sequence of 
reactions that are involved when $\hthreep$ reacts with CO:
\begin{eqnarray}
\hmp + \hm 		& \ra &		\hthreep + \h	\nonumber	\\
\hthreep + \co       & \ra & 	\hcop + \hm	\nonumber	\\
\el + \hmp 		& \ra &	\co + \h\,.		\nonumber 
\end{eqnarray}

The chemical heating due to the reactions of the $\hmp$ ion is then 
the rate at which that ion is produced per unit volume multiplied 
by the probability that it is transformed into an $\hthreep$ ion,  
(Eq.~\ref{convprob}), times the heating averaged over the three 
branches of the latter's reactions and including the heating 
from the dissociative recombination of $\hthreep$, all multiplied by the
$\hm$ fraction.
In terms of the notation of Eq.~\ref{chemQsum}, the chemical heating per ion pair 
stemming from the production of the $\hmp$ ion is, 
\be
\label{Qformula}
Q_{\rm chem}(\hmp) =  \frac{\xhm}{\xh + \xhm}\, F(\hmp) \, P(\hmp, \hthreep) \;  
[\Sigma_i \, B_i(\hmp) \, \overline{q}_i(\hmp) + B_{\rm e} \, \overline{q}_{\rm e}(\hmp) ]\; .
\ee
The heating energies $\overline{q}_i$ are given following Eq.~\ref{atomicOheat}, 
and the branchings are defined in Eq.~\ref{bratios}; $q_{\rm e}(\hmp) = 7.6$\,eV. 
For the case where ion-molecule destruction prevails over dissociative recombination, 
we can estimate the value of the chemical heating for the 
case where the abundances for CO, $\htwoo$, and O all equal $10^{-4}$. 
The result is $\Sigma_i \, B_i(\hmp) \, q_i(\hmp) = 8.4$\,eV, but this 
must still be multiplied by $F(\hmp)=0.88$ to yield $Q_{\rm chem}(\hmp) = 
7.2$\,eV.

Examination of Eqs.~\ref{waterheat} and \ref{atomicOheat} reveals that the products 
of the ion-molecule reactions for $\htwoo$ and O involve the 
radicals OH and H whose further reaction can lead to more 
chemical heating. For example, OH can be converted to $\htwoo$ by the 
exothermic reaction, 
\be
\label{ohreaction}
\oh + \hm \;\; \ra \;\; \htwoo + \h \,,
\ee
and the H atoms can make $\hm$ by formation on dust grain surfaces ``gr'',
\be
\label{h2form}
\h + \h + {\rm gr}  \;\; \ra \;\; \hm + {\rm gr}^\prime \,.
\ee
The energy yield of the OH reaction is 0.65\,eV. When this is added 
to the energy yields of Eqs.~\ref{waterheat} and \ref{atomicOheat}, the 
chemical heating produced in these reactions is increased slightly, 
$\overline{q}_2 = 8.4\,{\rm eV}$ and $\overline{q}_3 = 6.4\,{\rm eV}$. 
The previous estimate of $\Sigma_i \, B_i \, \overline{q}_i$ for the case of equal 
abundances for CO, $\htwoo$, and O (all $10^{-4}$) is increased from 8.4\,eV 
to 8.8\,eV, and  $Q_{\rm chem}(\hmp) = 7.7$\,eV.

The heating from Eq.~\ref{h2form} depends on how much if any of the energy 
release goes into kinetic energy of the newly formed molecule compared to internal 
excitation and excitation of the birth grain. Theoretical estimates and guesses 
in the literature range from 0.5 to 3.0\,eV, but laboratory experiments (e.g., 
Roser et al.~2003) suggest that the $\hm$ molecule thermally accommodates to 
the temperature of the grain before making its final escape, at least as far as 
kinetic energy is concerned. In a  recent experiment, Lemaire et al.~(2010) present 
evidence that $\sim 30$\% of newly formed $\hm$ molecules are vibrationally 
excited for dust temperatures as warm as 70\,K. Assuming that the results of Lemaire 
et al.~applies to molecular regions where the density is large enough for 
vibrationally excited $\hm$ molecules to be collisionally de-excited, the chemical 
heating for the $\htwoo$ and O channels would be increased by 1.5\,eV. 
This would then lead to an increase in the heating function by about 1\,eV 
to $Q_{\rm chem}(\hmp) \sim 9$\,eV. In the numerical estimates below, we ignore 
this contribution pending the resolution of this long standing issue regarding 
$\hm$ formation on grains.

\subsection{Chemical Heating of {\rm H}$^+$ and {\rm He}$^+$}

In principle, the chemical heating of $\hp$ and $\hep$ ions can be treated 
in a similar way as $\hmp$, but the outcomes depend on the chemical composition 
of the molecular region in question. Thus $\hp$ does not react 
with CO, the most abundant and stable heavy molecule, and its fast charge-exchange 
with O is closely balanced by the reverse reaction. It does react with 
$\htwoo$, which can be very abundant under certain circumstances, and with 
many organic species that usually have small abundances. In a 
cold molecular cloud, where $\htwoo$ may be expected to be frozen out on 
grains, this route will be shut down. On the other hand, in a dense warm 
region like those observed in the inner regions of protoplanetary disks, 
the primary destruction pathway for $\hp$ 
is   
\be 
\hp + \htwoo \;\; \ra \;\; \htwoop +\h \,,
\ee
followed by hydrogenation to $\hthreeop$,
\be 
\htwoop + \hm \;\; \ra \;\; \hthreeop +\h \,,
\ee
and destruction of $\hthreeop$ by dissociative recombination, according to 
Eq.~\ref{h3op+branches}. The energy balance equation is similar to 
Eq.~\ref{waterheat} for $\hmp$ reacting with $\htwoo$. The only difference 
is that $\hmp$ is replaced by $\hp$ on the left side and $\hm$ is replaced 
by H on the right side. The net {\it change} in exothermic reaction yield 
(and in the maximum heating) is just the difference between the ionization 
potentials of $\hm$ and $\h$, or 1.83\,eV. Thus the chemical heating arising 
from $\hp$ analogous to Eq.~\ref{Qformula} is, 
\be
\label{Qhp}
Q_{\rm chem}(\hp) =  \frac{\xhm}{\xh + \xhm}\; F(\hp)\;  B_2(\hp) \; 
\overline{q}_i(\hp) \,,
\ee
where $F(\hp)$ is the fraction of $\hp$ ions (0.040 from Table 1), 
$B_2(\hp) \simeq 1$ is the branching ratio for $\hp$ to react with $\htwoo$, 
and $\overline{q}_2(\hp) = \overline{q}_2(\hmp) - 1.83$\,eV. In principle, $B_2(\hp)$ 
is given by an expression similar to Eq.~\ref{bratios} with $\beta '$,
the dissociative recombination rate coefficient for $\hthreep$,  replaced by
the radiative recombination rate for $\hp$. Because the latter is smaller 
by many orders of magnitude, destruction of $\hp$ by radiative recombination 
can be ignored. For a completely molecular region, $F(\hp)=0.04$, $B_2 \sim 1$, 
and $\overline{q}_2(\hp) = 6.9$\,eV, resulting in $Q_{\rm chem}(\hp) = 0.28$\,eV. 
This estimate includes the energy that can be recovered from the OH radical 
but not from the conversion of product H atoms to form $\hm$ on grains. It 
is much smaller  than the heating per ion pair for $\hmp$ because $\hp$ is 
less energetic and because only 4\% of the ion pairs involve $\hp$. This 
estimate is also  sensitive to the physical conditions because the $\htwoo$ 
abundance can vary, especially due to freeze out on grains, and because 
reactions with other species may occur such as charge transfer with neutral 
heavy atoms.

The heating by the $\hep$ ions is more complicated because it reacts with 
many molecules including $\hm$. Here we consider the dominant reaction to be,
\be 
\hep + \co \;\; \ra \;\; \cp + {\rm  O} + \he \,,
\ee
followed by,
\be
\cp + \htwoo \;\; \ra \;\; \hcop +\h \,,
\ee
and then by dissociative recombination of $\hcop$. A similar analysis to that 
followed above for $\hmp$ gives an energy budget equation, 
\be
\label{hepheat}
\el + \hep + \htwoo \;\; \ra \;\; {\rm O} +  2\h + \he \,,
\ee
with an energy yield of 15.0\,eV. Including the reactions of the remaining O 
atom gives another 0.6\,eV, or $q_1(\hep)= 15.6$\,{\rm eV}. On this simplified 
basis, the heating due to $\hep$ ions is 
\be
\label{Qhep}
Q_{\rm chem}(\hep) =  \frac{\xhm}{\xh + \xhm}\; F(\hep)\; B_1(\hep) \; q_1(\hep) \,.
\ee
Using $F(\hep)=0.80$, $B_1(\hep) = 1$ (radiative recombination is again unimportant 
in molecular regions), $Q_{\rm chem}(\hep) = 1.25 $\,eV. This may be a slight overestimate 
because other reactions of $\hep$ have been neglected, e.g., with water.  

\subsection{Summary of Chemical Heating}

To summarize this treatment of chemical heating, we have considered the energy 
that is available for heating due to the reactions of $\hmp$, $\hp$, and $\hep$ 
ions, which account respectively for 88\%, 4\%, and 8\% of the $\sim 27$ ion pairs 
per keV produced by high energy electrons ($E > 500$\,eV). The $\hmp$ ion 
accounts for 7.7\,eV per ion pair, which is most of the chemical heating (84\%). 
This is very close to the heating from the dissociative recombination of 
$\hthreep$, $\simeq 7.6$~eV (GL73). Even the total chemical heating, 9.2\,eV 
per ion pair, is insensitive to the  electron fraction because the chemical 
heating from $\hp$ and $\hep$ are also essentially independent of the electron fraction. 
There are a number of fine points that could add or subtract 1-2 eV 
from this result, depending on the physical conditions and on the role of 
poorly understood processes. For example, this treatment of high-density 
chemical heating is an over-estimate in that no energy is assumed to be lost 
by vibrationally-excited molecules produced in the relevant chemical reactions, 
on the assumption that the density is high enough for collisional de-excitation 
to be effective. Similarly, we have ignored the uncertain possibility that 
$\sim 1.5$\,eV of excitation energy of newly formed $\hm$ molecules may be 
available for heating at high densities. 

\section{Results}
\label{results}

In the previous sections we developed a theory for the heating of molecular
gases exposed to cosmic rays and X-rays, with detailed consideration of 
chemical heating. To illustrate the magnitude of the various contributions 
to the heating, we list in Table 4 the maximum values for the case of zero 
electron fraction and densities high enough for vibrationally-excited 
molecules to be collisionally de-excited; the electron energy is $E = 1$~keV. 
The chemical heating in Table 4 is based on somewhat arbitrary choices for 
the most abundant neutrals: $\xco = \xo = \xhtwoo = 10^{-4}$.  Starting at 
the bottom, we see that the total maximum heating is $Q=18.7$~eV per ion pair; 
close to 50\% of the energy expended per ion pair, which in this case is 
$W = 36.8$~eV. The largest fraction, about one half, is accounted for by 
chemical heating (9.3\,eV). This value is $\sim 20 \%$ larger than the value 
of 7.6\,eV obtained from reactions Eq.~\ref{h2p}, 
$\hmp + \hm \rightarrow \hthreep + \h$, and 
Eq.~\ref{h3p_diss}, $\hthreep + \el \rightarrow \hm + \h$, along the lines 
of GL73. The next largest contribution, 5.2\,eV, comes from vibrationally 
excited $\hm$ molecules, produced by direct collisional excitation or 
following the excitation of the $B$ and $C$ states. Elastic scattering, rotational 
excitation, and dissociation account for the rest, or $\sim 20\%$.

\begin{deluxetable}{lrr}
\tablewidth{0pt}
\tablecaption{Heating per Ion Pair$^{\rm a}$}
\tablehead{
\colhead{Process}                         	& \colhead{$Q$~(eV)}
}
\startdata
$Q_{\rm el/rot}$                	& 2.1   &               \\ 
$Q_{\rm diss}$                  	& 2.1   &               \\
$Q_{\rm dir/vib}$ 			     	& 2.8   &               \\ 
$Q_{\rm BC/vib}$			  	& 2.4   &               \\
$Q_{\rm chem}(\hmp,\hm)$        	& 7.7   &               \\ 
$Q_{\rm chem}(\hp,\hm)$         	& 0.3   &               \\ 
$Q_{\rm chem}(\hmp,\hep)$       	& 1.3   &               \\ 
\tableline
Total heating $Q$               	& 18.7  &               \\ 
\enddata
\tablenotetext{a}{$E=1$\,keV}
\end{deluxetable}

The values in Table 4 are based on several assumptions: the electron 
fraction is very small, the density very high and the abundances of 
C, $\htwoo$ and O all equal to $10^{-4}$. The chemical heating 
appears to be insensitive to the electron fraction as long as 
Eq.~\ref{xel_condition} is satisfied. On the other hand, Coulomb 
collisions with ambient electrons begin to play a significant role 
in heating wherever $\xel$ approaches the $10^{-4}$ level. This 
effect is automatically included in Eq.~\ref{heating_efficiency}. 
The entries for vibrational heating in Table 4 are the most sensitive 
to density because very high densities are required to collisionally 
quench vibrationally excited $\hm$ molecules, as discussed in Section~2.4. 
This restriction means that vibrational heating is ineffective in 
molecular clouds outside of pre-stellar cores.  The chemical heating 
also depends weakly on the abundances of the neutral species.  For 
example, if instead of the abundances used for the estimates in Table 4, 
$\xco = \xo = \xhtwoo = 10^{-4}$, we choose 
$\xco = 10^{-4}$, $\xo = 2\times 10^{-4}$, and $\xhtwoo = 0$, 
the chemical heating from $\hmp$ increases by about $\sim 0.2$\,eV. 
Thus the chemical heating associated with the $\hmp$ ion is reasonably 
robust in molecular regions at $\sim 8-9$\,eV.

Under certain circumstances, such as very low energy cosmic-rays or soft 
X-rays, the considerations of this paper need to be extended to electron 
energies $E < 1$\,keV. Without attempting to treat this subject in detail, 
we can obtain a preliminary view of the situation from Table 5, which gives 
heating functions $Q$ and the energy to make an ion pair as a function of 
electron energy. Many of the heating functions increase with decreasing
energy, especially collisional excitation and dissociation. However the 
percentage of the energy to make an ion pair that goes into heating remains 
close to 50\% down to $E=50$\,eV. Indeed, the following quantities vary by
less than 10\% for energies greater than 200\,eV: total heating $Q$, energy to make an ion pair $W$ and the percentage heating.

\begin{deluxetable}{lrrrrrr}
\tablewidth{0pt}
\tablecaption{Heating $Q$ (in eV) per Ion Pair vs. Electron Energy}
\tablehead{
\colhead{Electron Energy $E$~(eV)}   & \colhead{30}   	& \colhead{50} 	& \colhead{100} & \colhead{200} 	& \colhead{500} & \colhead{1000}	
}
\startdata
$Q_{\rm el/rot}$		&  7.7 	&4.4	&3.0	&2.5	&2.1	& 2.1 	       \\
$Q_{\rm diss}$ 		&  12.0	&5.8	&3.9	&2.7	&2.3	& 2.1		\\
$Q_{\rm dir/vib}$		&  10.9 	&6.1	&4.0	&3.3	&2.9	& 2.8	 	\\
$Q_{\rm BC/vib}$		&  4.3 	&3.3	&2.9	&2.6	&2.4	& 2.4	 	\\
$Q_{\rm chem}(\hmp,\hm)$	&  8.6 	&8.3	&8.0	&7.8	&7.7	& 7.7		\\
$Q_{\rm chem}(\hp,\hm)$	&  0   	&0.1	&0.3	&0.3	&0.3	& 0.3	 	\\
$Q_{\rm chem}(\hmp,\hep)$	&  0.2 	&0.5	&0.8	&1.0	&1.2	& 1.3		\\
\tableline
Total heating $Q$		&  43.7	&28.5	&22.9	&20.2	&18.9	&18.7  	\\
\tableline
\tableline
Energy to make an ion pair $W$& 76.8 	&52.9 &43.1 	&39.6 	&37.4 	&36.8	\\
Number of ion pairs $N=E/W$       & 0.39 & 0.95 &  2.3 &  5.1 & 13.4 & 27.2 \\  
Percentage heating		    &57	&54	&53	&51	&51	&51			\\
\enddata
\end{deluxetable}

We can make the implications of the theory in this paper more concrete by 
estimating the heating for representative examples of interstellar and 
circumstellar matter that have significant amounts of $\hm$. The estimates 
in Table 6 are based on the assumption that the electron energy is greater 
than 0.5\,keV, corresponding to cosmic-ray proton energies greater than 1 MeV. 
For the well-observed line of sight towards $\zeta$\,Per ($A_{\rm V} = 1$) 
(an early focus of the $\hthreep$ observations, Indriolo et al.~2007), we 
use the modeling results for many observed species by Shaw et al.~(2008), 
who fit the cosmic-ray ionization rate per $\hm$ with 
$\zeta_2 \simeq 8 \times 10^{-16}\, \ps$. The densities in this diffuse 
cloud are too low for vibrational heating. The estimate for elastic/rotational 
heating may be too high because the density and temperature in this cloud may 
not be large enough for significant quenching of rotational excitation, according to the  discussion in Sec. 3.1. The chemical heating in this case comes from dissociative recombination because, according to Eq.~\ref{dr_dominance}, it dominates over 
ion reactions. All of the heating estimates for $\zeta$\,Per have been reduced 
by 70\% to take into account that atomic H has not been fully converted into 
$\hm$. The entries under $\hthreep$ destruction in Table 6 indicate whether the 
chemical heating is dominated by dissociative recombination of $\hthreep$
(``DR'') or ionic reactions (``I'').

For the case of a molecular cloud clump, dissociative recombination plays 
a role in the destruction of $\hthreep$, but ion-molecule reactions are 
probably more important. Again, clumps are not dense enough for vibrational heating, 
and they also may not be warm enough to quench rotational excitation. 
For the inner 
region of a pre-stellar core, the density may be high enough for vibrational heating, but their low temperatures suggest that the elastic/rotational 
heating may be somewhat smaller than given in Table 6. The freeze-out of 
volatiles, including some CO, means that the chemical heating will probably be close to the value for dissociative recombination.  

The densities in the inner molecular layer of a protoplanetary disk are often 
in excess of $10^9 \percc$, and both rotational and vibrational heating  
should be effective. Because these regions are close to a stellar X-ray source, 
the X-ray ionization rates are much larger than the cosmic-ray rates 
for interstellar matter. Thus the electron fractions are relatively
high at the top of the molecular region, $\sim 10^{-7} - 10^{-6}$, 
and Eq.~\ref{dr_dominance} indicates that dissociative recombination 
still plays a role in the chemical heating of the very dense inner 
regions. However, for sufficiently large vertical columns, the X-ray  
ionization is sufficiently reduced that ionic reactions dominate chemical 
heating.

\begin{deluxetable}{lcccc}
\tablewidth{0pt}
\tablecaption{X-ray and Cosmic-ray Heating (in eV) in Molecular Regions}
\tablehead{
					& \colhead{$\zeta$\,Per} 	& \colhead{molecular cloud}		& \colhead{prestellar core} 	& \colhead{protoplanetary disk}	 \\ 
					& \colhead{diffuse cloud}	& \colhead{clump} 			& \colhead{inner region} 	& \colhead{active region at 1\,AU}	
}       
\startdata
$n_{\rm H}$~(cm$^{-3}$)         &80      	&  $10^{4}$     & $10^{7}$ 	      & $10^{10}$\\
$T$ (K)                         &$\simeq 60$	&  10  		&    6     	      & 1000	\\ 
$\xel$                          & $2 \times 10^{-4}$  & $10^{-7}$ & $10^{-9}$ & $10^{-6}$	\\
\tableline
$\hthreep$ destruction		& DR$^*$    	& DR + I$^*$	& DR + I              & DR + I 	\\
\tableline
$Q_{\rm el/rot}$~(eV)          	& 4	   	& 2   		& 2		      & 2 	\\
$Q_{\rm vib}$~(eV)	       	& 0	        & 0    		& 5	              & 5	\\
$Q_{\rm diss}$~(eV) 	        & 1		& 2    		& 2    	              & 2	\\
$Q_{\rm chem}$~(eV) 	     	& 5		& 9 	 	& 8                   & 9       \\
\tableline
Total heating $Q$~(eV)          	       	& 10 	        & 13   		& 17  	              & 18      \enddata
\tablenotetext{*}{DR stands for dissociative recombination and I for ionic reactions.}
\end{deluxetable}

All of the cases considered in Table~6 contain a significant if not an
overwhelming fraction of $\hm$. These rough estimates for the total
heating per ion pair increase with the total density, and range from  
$\sim 10$\,eV per ion pair for a moderately thick diffuse cloud like 
$\zeta$\,Per to  $\sim 18$\,eV per ion pair for the densest regions.
Perhaps the most important conclusion from these four examples is that
cosmic-ray and X-ray heating are sensitive to the physical conditions, 
as expressed in the equations presented in Sections 3 and 4. The values of $Q$ 
listed in Table~6 for various environments, combined with prescriptions 
for $\zeta_{\rm H}$ as function of the column density of the ambient gas 
(Padovani et al.~2009) and for the magnetic field intensity 
(Padovani \& Galli~2011), allow the determination of the cosmic-ray heating rate 
in clouds, cores, and protostellar disks.

\section{Conclusion}
\label{conc}

Dalgarno et al.~(1999; DYL) made an extensive study of the energy 
loss of fast electrons in H, He and $\hm$, He gas mixtures. The electrons 
are at the heart of the interaction of cosmic rays and X-rays with 
interstellar and circumstellar matter. Although DYL analyzed essentially 
all excitation and ionization processes, they only discussed the heating 
from elastic collisions and rotational excitation of $\hm$. Starting from 
the results in DYL, we have extended the scope of cosmic-ray and X-ray 
heating to include all of the relevant interactions. One of the main 
conclusions of this study is that heating by fast-electrons depends on 
the physical properties of the gas, i.e., on the abundance of electrons, 
$\hm$ molecules, and heavy atoms and molecules, and also on the total 
density of hydrogen nuclei. The electron fraction is important because, 
once it exceeds a certain level, heating by collisions with ambient 
electrons becomes important. The electron abundance also determines 
whether the destruction of the $\hthreep$ ions proceeds by dissociative 
recombination or ionic reactions, which affects the quantitative amount
of chemical heating. Of course the $\hm$/H abundance ratio is important 
because the diversity of the energy levels of $\hm$ offers many more channels for 
energy loss than atomic H. The dependence on physical conditions means 
that X-ray and cosmic-ray heating cannot be specified by a single number. 
This should be clear from the fact that in neutral atomic regions the 
heating efficiency is only 12\%, whereas in neutral molecular regions 
it can reach 50\% at very high densities.

A wide range of values for the heating per ion pair can be found in the
literature on molecular clouds, from 7~eV (Stahler \& Palla 2004), to
20~eV (Goldsmith \& Langer 1978, Goldsmith~2001), but not necessarily
for the reasons given here, e.g., in the above discussion of Table 6.
Many authors adopt Goldsmith's value while others estimate intermediate
values of 10--15~eV based on GL73, e.g., Maloney et al.~(1996),
Yusef-Zadeh et al.~(2007), and Krumholz et al.~(2011). As shown in Table
4, 13~eV is a good choice for not too dense molecular gas.

Chemical heating also applies to regions exposed to far UV radiation 
(Dalgarno and Oppenheimer 1974). It was then considered by 
Barsuhn \& Walmsley~(1977) and Clavel et al.~(1978), who studied the 
chemical and thermal equilibrium in dark clouds exposed to far UV 
radiation and cosmic rays. Clavel et al. explicitly included the 
contribution of every reaction in their chemical network to heating. 
Chemical heating has also been widely used in the study of planetary 
atmospheres (e.g., Roble et al.\,1987). 

The results of this paper are based on the fact that roughly half of the 
energy generated by cosmic rays and X-rays comes goes into ionization of the 
gas and roughy half into its excitation, more specifically according to the way  
the individual processes are treated by DYL. Roughly half of the gas 
heating comes from the reactions of various ions with the assumed mainly 
neutral gas. Potentially an equal amount of heating can arise from 
dissociation and from rotational and vibrational excitation, but the yield from excitation depends on whether the physical conditions are conducive to the 
quenching of the excited levels. If quenching is inefficient, the levels decay with the emission of radiation, which can escape or be absorbed by dust. This possibility is even more important for the excitation of the singlet levels of 
$\hm$, e.g., the $B$ and $C$ levels, and the excitation of the $n=2$ level of H in atomic regions. We have not attempted to follow the fluorescent radiation to determine how much escapes and how much is absorbed. This is an important issue when considering the broader thermal properties of the gas. The treatment of the fluorescent radiation involves radiation transfer and depends on the properties of the dust, i.e., it is application specific. In contrast, our goal has been to treat one relatively well-defined part of the thermal problem of interstellar and circumstellar molecular gas.  

In pursuing this goal, we have also ignored the direct interaction of X-rays and cosmic-rays with the dust, of some interest because of the possibility that they may release fast electrons from the dust. Although this occurs, it is relatively unimportant. First of all, the dust cross section per H nucleus for the MRN distribution for the diffuse ISM is $1.6\times 10^{-21} \psqcm$. For molecular clouds and disk atmospheres, it will be even smaller. The main inelastic cross sections for a keV electron with $\hm$ are typically several times 
$10^{-17} \psqcm$, so the probability that an incident X-ray or cosmic ray interacts with dust is roughly 1000 time less than with $\hm$. And when this rare event does occur, no more than 1\% goes into in photoelectrons, as shown by Dwek and Smith (1996) for the EUV/Xray bands.

\section{Acknowledgements}

We are indebted to M\'{a}t\'{e}  \'{A}d\'{a}mkovics for testing the 
heating terms described in this paper in a thermal-chemical code 
for protoplanetary disks. One of us (A.E.G.) acknowledges support 
from NASA grant NNG06GF88G (Origins) and NASA grant 1367693 
({\it Herschel} DIGIT). The authors thank an anonymous referee for 
a careful reading and thoughtful comments.

\appendix
\section{Chemical Heating in Atomic Regions}

Table A provides the same information on the energy partitions for an atomic H, He mixture as Table 1 does for a $\hm$, He mixture. Although less energy is expended in ion production (because of the smaller ionization potential of H compared to that for $\hm$), the total number of ions is essentially unchanged at 27.2 ion pairs per keV.  
\begin{center}
\begin{tabular}{lcccc}   
\multicolumn{5}{c}{Table A. DYL Energy Partitions for a He, $\h$ mixture, 
$\xel = 0$ and $E = 1$\,keV} \\
\hline\hline
Process & $W$	& 1\,keV/$W$ & Specific energy 	& Total energy   \\ 
       & (eV) & & (eV) & (eV) \\      
\hline
$\hp$			& 39.8	 & 25.1	& 13.60	& 342	\\
$\hep$			& 487	 & 2.05	& 24.6		&  51	\\
He$^{2+}$		& 16400 & 0.061	& 54.4		&   3	\\
\hline
Total Ion production	&	 &		&		& 396	\\
\hline
H\,2$^1$S		& 267	 & 3.75	& 10.2		&  38	\\
H\,2$^1$P		& 33.1	 & 30.2	& 10.2		& 308	\\
H\,$n=3$		& 191	 & 5.24	& 12.1		&   6	\\
H\,$n>3$		& 236	 & 4.24	& 13.2		&  56	\\
\hline
Total $\h$ excitation	&	 &		&		& 408	\\
\hline
He\,2$^1$S		& 17300 & 0.058	& 20.6		&   1.2		\\
He\,2$^1$P		& 2080	 & 0.481	& 21.2		&  10.2		\\
He\,2$^3$S		& 50700 & 0.020	& 19.8		&   0.39	\\
He\,2$^3$P+$n>2^3$	& 30800 & 0.032	& 21.0		&   0.68	\\
He\,$n>2^1$		& 3790	  & 0.264	& 23		&   6.1		\\
\hline
Total He excitation		&	 &		&	&  19	\\
\hline				
Total excitation	&	 &		&		& 427	\\
\hline	
Heating			&	 &		&	& 117	\\
\hline
\hline
Grand total		&	 &		&		& 940 \\
\hline
\end{tabular}
\end{center}

The heating of an atomic region is simpler to treat than a molecular
region because of the absence of rotational and vibrational
excitation.  The direct heating by elastic scattering is usually
important.  We ignore the radiation emitted following the electronic
excitation of the levels of H and He in Table A, and focus here on the
chemical heating.

In addition to radiative recombination,
\be
\el + \hp \, \ra \; \h + h\nu, \hspace{.5in} 
\alpha(\hp) \simeq 2.12 \times 10^{-10}T^{-0.73} \ccps\,,
\ee
$\hp$ can be destroyed by radiative attachment to H to form $\hmp$,
\be
\label{h+hreac}
\hp + \h \, \ra \; \hmp + h\nu, \hspace{.5in} 
k_{\rm ra} \simeq 2.10 \times 10^{-23}\, T^{1.80} \ccps\,.
\ee
The rate coefficient for radiative association is especially small, and
the probability to form $\hmp$ this way is given by,
\be
\label{probh+toh2+}
P(\hp,\hmp;\h) = \frac{k_{\rm ra}\xh}{k_{\rm ra}\xh+ \alpha \xel} \,
                     = \frac{\xh}{\xh + (\alpha / k_{\rm ra}) \xel}\,.
\ee
In order for radiative recombination to be important, the electron fraction must 
approach the value,
\be
\frac{k_{\rm ra}}{\alpha} \simeq 10^{-13}\, T^{2.53}\,.
\ee
This requirement is relatively easy to satisfy, e.g., 
$k_{\rm ra} / \alpha \sim 10^{-8}$ at 100\,K and $\sim10^{-4}$ at 4000\,K. 

The energetics of the two processes are different. In both cases the 
radiation is lost or absorbed by dust grains and will be ignored.
In an atomic H region the newly formed $\hmp$ can charge-exchange with H 
to form $\hm$,
\be
\label{h+chex}
\hmp + \h \,\ra \; \hm + \hp\,,  \hspace{0.5in}
k = 6.4 \times 10^{-10} \ccps\,.
\ee 
This reaction can provide up to $q(\hp;\h) = IP(\hm) - IP(\hmp)=1.83$\,eV in heating, where $IP$ stands for ionization potential, but it has to compete with dissociative recombination, Eq.~\ref{h2p}, which has a branching ratio,
\be
B_{\rm e}= \frac{\beta \xel}{\beta \xel + k \xh } 
         =  \frac{(\beta / k) \xel}{(\beta / k) \xel + \xh }\,.
\ee
The ratio $k / \beta = 5.33 \times 10^{-3}\, T^{0.4}$ 
determines the electron fraction where dissociative recombination is important,
i.e., in regions where $\xel \sim 0.01$. 

We follow Section 3 and express the heating per ion pair for atomic regions 
due to the $\hp$ ion as,  
\be
\label{Qhpatomic}
Q_{\rm chem}(\hp;\h) =  \frac{\xh}{\xh + \xhm}\; F(\hp;\h)\, 
P(\hp,\hmp;\h)\, 
[B_{\h}\, q_{\h}(\hp;\h) +  B_{\rm e}\,q_{\rm e}(\hp;\h)] \,,
\ee
where $F(\hp;\h)$ is the fraction of all ions that are $\hp$; $P(\hp,\hmp;\h)$ 
is the probability that an $\hp$ ion is converted into $\hmp$; 
$q_{\h}(\hp;\h) = 1.83$\,eV is the net heating from reactions Eqs.~\ref{h+hreac} 
and ~\ref{h+chex}; $B_{\h}$ is the branching ratio for reaction Eq.~\ref{h+chex}, 
\be
B_{\h}= \frac{k \xh}{k \xh  + \beta \xel } 
         =  \frac{k \xh}{\xh + (\beta / k) \xel }\,
\ee
and $q_{\rm e}(\hp;\h)=11.0$\,eV is the net heating from radiative charge exchange,  Eq.~\ref{h+hreac}, followed by dissociative recombination, Eq.~\ref{h2p}. From Table A, we find that the fraction of $\hp$ ions 
$F(\hp;\h)= 0.92$ for the parameters in that table ($E=1$\,keV and $\xel = 0$); it increases slowly with decreasing electron energy.

The maximum chemical heating per ion pair coming from $\hp$ depends on 
how the $\hmp$ it generates is destroyed: 1.7\,eV by charge exchange and 10.1\,eV 
by dissociative recombination, with the latter only occurring for relatively large electron fractions. However in most H\,I regions, the dominant factor is the
probability of forming $\hp$, $P(\hp,\hmp;\h)$ in Eq.~\ref{Qhpatomic}, which can greatly reduce $Q_{\rm chem}(\hp;\h)$, simply because radiative association 
in this case is such a weak process. Thus chemical heating in atomic H regions is usually negligible.

In addition to radiative recombination,
\be
\el + \hep \, \ra \; \he + h\nu\,, \hspace{.5in} 
\alpha(\hep) \simeq 2.12 \times 10^{-10}T^{-0.73} \ccps\,,
\ee
$\hep$ can be destroyed by radiative charge exchange,
\be
\label{hepradchex}
\hep + \h \, \ra \; \hp + \he + h\nu\,, \hspace{.5in} 
k_{\rm chex} \simeq 1.6 \times 10^{-16}\, T^{0.50} \ccps\,,
\ee
and by radiative association to form $\hehp$,
\be
\label{hepradassn}
\hep + \h \, \ra \; \hehp + h\nu\,, \hspace{.5in} 
k_{\rm ra} \simeq 1.33 \times 10^{-14}\, T^{-0.37} \ccps\,.
\ee
All of these radiative processes are weak, although the last two are not as weak 
as the corresponding reactions just discussed that start from $\hp + \he$. These 
reactions have the potential to provide greater heating because of the larger 
energy of the $\hep$ ion, except that much of the available energy is lost in radiation.  We therefore ignore radiative recombination and focus on 
radiative charge exchange, Eq.~\ref{hepradchex}, and radiative association, 
Eq.~\ref{hepradassn}. The former process generates $IP(\he) - IP(\h)=11.0$\,eV in recombination radiation rather than gas heating. The heating from the formation of the $\hehp$ ion, discussed in the previous section, has a {\it maximum} value 1.83\,eV.  But this value must be reduced by the fraction of $\hep$ which, according to Table A for $E=1$\,keV, is $F(\hep,\h)= 0.075$. It may also be reduced by the branching ratio for radiative association, 
\be
B_{\rm ra}(\hep) = \frac{k_{\rm ra}}{k_{\rm ra}+ k_{\rm chex} + \alpha \xel}\,,
\ee 
which is often less than unity. For example, in warm regions charge exchange is more important than radiative association, and radiative recombination is competitive with the other reactions for
$\xel > 10^{-3}$. Thus we cannot expect the chemical heating from $\hep$ to be any more than 0.14\,eV, and thus negligible, as we found also for the $\hp$ ion. 


\begin{thebibliography}{}

\bibitem{a97}
Abgrall, H., Roueff, E., Liu, X., Shemansky, D.~E.\ 1997, ApJ, 481, 557 

\bibitem{bal97}
Balakrishnan, N., Forrey, R. C.,  Dalgarno, A. 1999, ApJ, 514, 520

\bibitem{bw77}
Barsuhn, J., Walmsley, C. M.\ 1977, A\&A, 54, 345

\bibitem{btes53}
Bates, D. R. \& Griffing, G.\ 1953, Proc. Phys. Soc., A66, 961 

\bibitem{c78}
Clavel, J., Viala, Y.~P., Bel, N.\ 1978, A\&A, 65, 435 

\bibitem{c07}
Crapsi, A., Caselli, P., Walmsley, M.~C., Tafalla, M.\ 2007, A\&A, 470, 221 

\bibitem{cvd75}
Cravens, T.~E., Victor, G.~A., Dalgarno, A.\ 1975, P\&SS, 23, 1059 

\bibitem{cd78}
Cravens, T.~E., Dalgarno, A.\ 1978, ApJ, 219, 750 

\bibitem{do74}
Dalgarno, A., Oppenheimer, M.\ 1974, ApJ, 192, 597 

\bibitem{dyl}
Dalgarno, A., Yan, M., Liu, W.\ 1999, ApJS, 125, 237 (DYL)

\bibitem{dw96}
Dwek, E., Smith, R. K. 1996, ApJ, 459, 686

\bibitem{ez11}
Everett, J. E., Zweibel, E. G. 2011, ApJ 738, 60 

\bibitem{g02}
Galli, D., Walmsley, M., Gon{\c c}alves, J.\ 2002, A\&A, 394, 275 

\bibitem{gl73}
Glassgold, A.~E., Langer, W.~D.\ 1973, ApJ 186, 859 (GL73)

\bibitem{gl78}
Goldsmith, P.~F.,Langer, W.~D.\ 1978, ApJ , 222, 881

\bibitem{g01}
Goldsmith, P.~F.\ 2001, ApJ 557, 736

\bibitem{hme69}
Henry, R.~J.~W., McElroy, M.~B.\ 1969, J. Atm. Sci., 26, 912 

\bibitem{hoop61}
Hooper, J. W., McDaniel, E. W., Martin, D. W., Harmer, D. S. 1961. 
Phys. Rev., 131, 1123

\bibitem{in07}
Indriolo, N., Geballe, T. R., Oka, T, McCall, B. J. 2007, ApJ, 671, 1736

\bibitem{in09}
Indriolo, N., Fields, B. D., McCall, B. J. 2009, ApJ, 694, 257

\bibitem{in12}
Indriolo, N., McCall, B. J. 2012, ApJ, 745, 911

\bibitem{j07} 
Jasche, J., Ciardi, B., En{\ss}lin, T.~A.\ 2007, MNRAS, 380, 417 

\bibitem{k07}
Krumholz, M.~R., Leroy, A.~K., McKee, C.~F.\ 2011, ApJ, 731, 25 

\bibitem{leb99}
Le Bourlot, J, Pineau des For\^{e}ts \& Flower, D. R.~1999, MNRAS, 305, 802 


\bibitem{lee08}
Lee, T.-G., Balakrishnan, N., Forrey, R. C., Stancil, P. C., Shaw, G., Schultz, D. R., Ferland, G. J. 2008, ApJ, 689, 1105

\bibitem{mht96}
Maloney, P.~R., Hollenbach, D.~J., Tielens, A.~G.~G.~M.\ 1996, ApJ, 466, 561

\bibitem{opal71}
Opal, C. B., Peterson, W. K. \& Beatty, E. C. 1971, J. Chem. Phys, 55, 4100  

\bibitem{p07}
Pagani, L., Bacmann, A., Cabrit, S., Vastel, C.\ 2007, A\&A, 467, 179 

\bibitem{p10}
Padovani, M., Galli, D., Glassgold, A.~E.\ 2009, A\&A, 501, 619
 
\bibitem{pg11}
Padovani, M., Galli, D.\ 2011, A\&A, 530, A109 

\bibitem{pb07}
Pineda, J.~L., Bensch, F.\ 2007, A\&A, 470, 615 

\bibitem{r12}
Rimmer, P. B., Herbst, E., Morata, O., Roueff, E. 2012, A\&A, 537, A7

\bibitem{r87}
Roble, R.~G., Ridley, E.~C., Dickinson, R.~E.\ 1987, J. Geophys. Res., 92, 8745 

\bibitem{r03}
Roser, J.~E., Swords, S., Vidali, G., Manic{\`o}, G., Pirronello, V.\ 2003, ApJL, 596, L55 

\bibitem{s94}
Schneider, I. F., Dulieu, O., Giusti-Suzor, A., Roueff, E.\ 1994, ApJ, 424, 983 (errata in ApJ, 486, 580)

\bibitem{sh08}
Shaw, G., Ferland, G.~J., Srianand, R., et al.\ 2008, ApJ, 675, 405 

\bibitem{s09} 
Shaw, G., Ferland, G.~J., Henney, W.~J., et al.\ 2009, ApJ, 701, 677 

\bibitem{sp04}
Stahler, S. W., Palla, F. 2004, Sec. 7.2.1, ``The Formation of Stars''.
(Wiley-VCH; Weinheim)

\bibitem{s93} 
Suchkov, A., Allen, R.~J., Heckman, T.~M.\ 1993, ApJ, 413, 542 

\bibitem{su94}
Sundstr\"om, G. Mowat, J. R., Danared, H. et al.\ 1994, Science, 263, 785

\bibitem{th74}
Theard, L. P., Huntress, W. T.\ 1974, J. Chem. Phys., 60, 2840

\bibitem{t00}
Tielens, A. G. G. M. 2000, in ``The Physics and Chemistry of the Interstellar Medium'' (Cambridge) 

\bibitem{t97}
Tin\'e, S., Lepp, S., Gredel, R., Dalgarno, A.\ 1997, ApJ, 481, 282

\bibitem{wr07}
Wrathmall, S. A., Gusdorf, A, Fower, D.R. 2011, MNRAS, 382, 133

\bibitem{yz07}
Yusef-Zadeh, F., Wardle, M., Roy, S.\ 2007, ApJL, 665, L123 

\end{thebibliography}
\end{document}